\documentclass[useAMS,usenatbib,usegraphicx]{mn2e}
\usepackage{ulem} 
\usepackage[dvips]{graphicx,color}
\usepackage{dcolumn}
\usepackage{bm}
\newcommand{\bvector}[1]{\mbox{\boldmath $#1$}}

\def\aap{A\&A}

\def\apj{ApJ}

\def\pasp{PASP}

\title[Novel formulation for rotational equilibria]
{A Novel Formulation by Lagrangian Variational Principle \\
for Rotational Equilibria: \\
Toward Multi-Dimensional Stellar Evolutions}

\author[N.Yasutake, K.Fujisawa, \&  S. Yamada]
       {Nobutoshi Yasutake$^1$,
       Kotaro Fujisawa$^2$,
       Shoichi Yamada$^2$\\
       $^1$Physics Department, Chiba Institute of Technology, 
       Shibazono 2-1-1, Narashino, Chiba, 275-0023, Japan\\
       $^2$Advanced Research Institute for Science and Engineering, Waseda University, 
       Okubo 3-4-1, Shinjuku, Tokyo, 169-8555, Japan
}

\begin{document}

\maketitle

\begin{abstract}
We have developed a new formulation to obtain self-gravitating, axisymmetric configurations in permanent rotation. The formulation is based on the Lagrangian variational principle, and treats not only barotropic but also baroclinic equations of state, for which angular momentum distributions are not necessarily cylindrical. We adopt a Monte Carlo technique, which is analogous to those employed in other fields, e.g.  nuclear physics, in minimizing the energy functional, which is evaluated on a triangulated mesh. This letter is a proof of principle and detailed comparisons with existing results will be reported in the sequel, but  some test calculations are presented, in which we have achieved an error of $O(10^{-4})$ in the Virial relation. We have in mind the application of this method to two-dimensional calculations of the evolutions of rotating stars, for which the Lagrangian formulation is best suited.
\end{abstract}


\section{INTRODUCTION}
After the invention of the Henyey method~\citep{henyey64}, our understanding of 
the structures and evolutions of spherically symmetric stars in the absence of rotation (and/or magnetic fields) have been advanced exponentially over the years by many researchers\citep{maeder00}. Although the stellar evolution is a time-dependent problem, it is actually so slow in most stages compared with the dynamical time scale that it is well approximated by a collection of time-independent configurations in equilibrium with gradually changing parameters such as the central density and temperature as well as chemical compositions. 

In the presence of rotation, however, the same method cannot be employed. One of the difficulties in calculating rotational equilibria is the fact that we do not know a priori what the distribution of angular momentum looks like in the stellar interior. This is indeed the case even for barotropes, in which matter pressure is a function of density alone, although it is mathematically shown in this case that the rotational equilibria have cylindrical distributions of specific angular momentum. 
When the angular momentum takes constant values on concentric cylinders, there exists a first integral and there are a variety of numerical treatments~\citep{ostriker, hachisu86}.
The realistic equation of state, however, is not barotropic but baroclinic, in which pressure depends on density, entropy and chemical compositions. 
It is hence a non-trivial issue how to make an appropriate sequence out of an ensemble of rotational configurations even if they are somehow obtained~\citep{uryu94, espinoza13}.  As a matter of fact, in the application of rotational equilibria, we would then need to know the temporal evolutions of thermodynamics quantities as well as angular momentum distribution, which would not be easy in multi-dimensions, since fluid elements composing rotating stars could change their positions non-radially in complicated ways even without convection. Such gradual evolutions that occur on the secular time scale are best treated by the Lagrangian formulation just as in the spherically symmetric case. This is the idea underlying our method.

It is observationally well known that massive stars rotate at a surface velocity of $\sim 200\ {\rm km\ s}^{-1}$ on the main sequence~\citep{fukuda82}. Although the angular momentum in  stellar interior is largely unknown, it will not be a small value. In fact, centrifugal force by such rotations can play a crucial role in the advanced evolutionary stages of massive stars~\citep{endal76}. In particular, various (magneto-) hydrodynamical instabilities induced by rotation have been intensively and extensively investigated over the years by many researchers (see e.g. \citealt{maeder00, maeder13}). Full-fledged instabilities may produce turbulence even in the radiative zones~\citep{mathis13}. 
Since such turbulences are expected to be anisotropic in general, the time-averaged states will be approximated by some baroclinic rotational equilibria. 

In this letter, we propose a novel idea to obtain rotational equilibria numerically for both barotropes and baroclines\footnote{Non-barotropic equations of state are collectively referred 
to as baroclinic equations of state.}, which is well suited for the application to the stellar evolution calculations. It is based on the Lagrangian variational principle: 
an energy functional is defined in such a way that its stationary point should correspond to a rotational equilibrium. It is then approximated on a triangulated mesh and 
minimized by moving grid points, a Monte-Carlo technique quite analogous to what is commonly used in other fields, e.g. in nuclear physics to obtain deformed nuclei. We 
demonstrate its nice performance in some test models although this letter is meant to be a proof of principle and detailed validations will be presented in the sequel. 

\section{FORMULATION}

\subsection{Lagrangian variational principle}

Our formulation is based on the Lagrangian variational principle, in which rotational configurations are obtained by minimizing the 
following functional under the conditions stated below:
\begin{equation}
E[\bvector{\xi}(\bvector{r})] = \int\!\! \varepsilon \rho \, dV + \frac{1}{2} \! \int \!\! \rho \phi \, dV + \int \!\! \frac{1}{2} \, \rho \! \left ( \frac{j}{\varpi} \right )^2 dV. 
\end{equation} 
In this expression, $\varpi$ is the distance from the rotation axis and $dV$ stands for the volume element; the integration domain is the stellar interior; $\rho$, $\varepsilon$, 
$\phi$ and $j$ denote the density, specific internal energy, gravitational potential and specific angular momentum, respectively. They are regarded as functionals of the 
so-called Lagrangian displacement vector $\bvector{\xi}(\bvector{r}) \equiv \bvector{r}'(\bvector{r}) - \bvector{r}$, which connects the fluid element at $\bvector{r}$ in an 
arbitrarily chosen reference configuration to the corresponding fluid element shifted to $\bvector{r}'$ in the configuration obtained by the deformation of the reference. Then a 
rotational equilibrium is obtained from a deformation or $\bvector{\xi}$ that corresponds to a stationary point of the functional $E[\bvector{\xi}]$. In fact, the Euler-Lagrange 
derivative of the functional gives the equations for rotational equilibrium:  
\begin{equation}
\frac{\delta E[\bvector{\xi}] }{ \delta \bvector{\xi}} 
= \nabla P + \rho \nabla \phi - \frac{\rho j^2}{\varpi^3} {\bf e_{\varpi}}
= 0, 
\label{eq:equilibrium}
\end{equation}
where $P$ and ${\bf e_{\varpi}}$ are pressure and the unit vector perpendicular to the rotation axis. In taking the Euler-Lagrange derivative in Eq.(\ref{eq:equilibrium}), 
the specific entropy (denoted by $s$ hereafter) and angular momentum are fixed for each fluid element, which is why this formulation is referred to as {\it Lagrangian}.
 
Note that the Lagrangian displacement vectors themselves are not necessarily small. The density as a functional of $\bvector{\xi}$ is given by mass conservation: 
$\rho(\bvector{r}') = J(\bvector{r}) \rho_{0}(\bvector{r})$. Here $\rho_0(\bvector{r})$ is the density distribution in the reference configuration and $J(r)$ is the Jacobian at $\bvector{r}$ for the 
transformation $\bvector{r} \rightarrow \bvector{r}' = \bvector{r} + \bvector{\xi}(\bvector{r})$. Since the specific internal energy is a function of density and specific entropy and the 
latter is unchanged by deformation, $\varepsilon$ can be regarded as a functional of $\bvector{\xi}$. The gravitational potential is obtained from the density by the Poisson equation 
\begin{equation}
\Delta \phi = 4 \pi G \rho,
\label{eq:poisson}
\end{equation}
where $G$ is the gravitational constant. In this letter, we solve Eq.(\ref{eq:poisson}) for a given density distribution and consider the gravitational potential also as a functional 
of $\bvector{\xi}$. The gravitational potential can be treated as an independent variational variable. The energy functional 
should be modified then to 
\begin{eqnarray}
E[\bvector{\xi}(\bvector{r})] & = & \int\!\! \varepsilon \rho \, dV + \int \!\! \rho \phi \, dV  +  \int \!\! \frac{1}{2} \, \rho \! \left ( \frac{j}{\varpi} \right )^2 \! dV \nonumber \\
&& + \frac{1}{4 \pi G} \int \frac{1}{2} (\nabla \phi)^2 dV.
\end{eqnarray} 
In this case the Poisson equation is obtained as a result of the Euler-Lagrange derivative with respective to $\phi$. This possibility will be detailed in our forthcoming paper. 

The Lagrangian variational principle presented above is not our own invention but has been known for many years \citep{tassoul78}. The following implementation, however, 
is our original idea. We adopt a finite element method: the meridian section of an axisymmetric star is covered by a triangulated mesh, and its grid points or nodes are moved to 
artificially deform the star; all physical quantities are assigned to the nodes; in particular, the mass, specific entropy and angular momentum allotted to each node are fixed in the 
node shifts; the energy functional then becomes approximately a function of the coordinates of nodes and its minimization gives a rotational equilibrium for given distributions of
mass, specific entropy and angular momentum on the triangulated mesh. It is emphasized that the variables to be solved in this formulation are the coordinates of each grid point 
unlike in the common Eulerian formulations, in which values of physical quantities on fixed grid points are seeked. 

The approximate energy functional is given as follows:
\begin{equation}
E_{\rm FEM}(\bvector{r}_i) = \sum_{i}\varepsilon _i m_i + \frac{1}{2}\sum_{i}\phi _i m_i + \sum_{i}\frac{1}{2} \!\left( \frac{j_i}{\varpi _i} \right )^2 \! m_i,
\label{eq:eq2}
\end{equation}
which is now actually a function of the coordinates of all nodes, $\bvector{r}_i$, with the subscript specifying the individual node. In this expression $m_i$ denotes the mass
assigned to the $i$-th node, which is also constant just like the specific entropy and angular momentum. The density at the $i$-th node is expressed as $\rho_i = m_i / V_i$, 
in which $V_i$ is a volume allotted to the $i$-th node and evaluated as a function of the coordinates of nodes from the volumes of the cells attached to this node. This 
corresponds to an approximate evaluation of the Jacobian $J(\bvector{r}_i)$ at the location of the $i$-th node.

It should be then obvious that the specific internal energy $\varepsilon _i$ is also a function of the coordinates of nodes, since the specific entropy $s_i$ is fixed. As for the 
gravitational potential $\phi _i$, the Poisson equation, Eq.(\ref{eq:poisson}), is solved numerically on the same triangulated mesh for the above $\rho_i$, which makes it possible for us to 
regard $\phi _i$ as a function of node positions again. With the energy function at hand this way, all we need to do is to minimize it with respect to the node coordinates for a given triplets of 
($m_i$, $s_i$, $j_i$).

\subsection{Minimization}

There are two possibilities to obtain a rotational equilibrium that corresponds to the minimum of the energy function given by Eq.~(\ref{eq:eq2}). The first choice is that one takes 
its derivatives with respect to $\bvector{r}_i$ to generate a system of coupled nonlinear equations for $\bvector{r}_i$ and solve it numerically. The resultant equations can be
regarded as an approximate expression of the force balance among gravity, pressure and centrifugal force. The second option, a Monte Carlo technique, is to randomly shuffle 
the coordinates of the grid positions and search for the configuration that gives the minimum to the energy function.
In our formulation we choose the latter. 
 
There a couple of reasons. It is known that different Lagrange displacements may give the same Eulerian variations in some cases~\citep{friedman}. An example is the isentropic case, i.e. 
the specific entropy is uniform in the stellar interior. This is most easily understood from the fact  that the infinitesimal Lagrangian displacement vector $\bvector{\xi}$ that assumes 
the form of $\bvector{\xi} = 1/\rho \nabla \times \bvector{\eta}$ for an arbitrary axisymmetric vector field $\bvector{\eta}$ does not change density as a function of spatial position. Such a 
{\it gauge freedom} renders the system of equations for $\bvector{r}_i$ indefinite and would make the matrix obtained by linearization singular if one were to employ the Newton-Raphson method. 
This means that, without an appropriate gauge fixing, the formulation could not be applied to isentropic configurations, which are normally supposed to be the simplest case. Even if the 
specific entropy is not homogeneous, the energy function is rather insensitive to the displacements of nodes close to the rotation axis and/or to the surface in general, which implies that there are 
always small (i.e. almost singular) eigenvalues in the matrix of the linearized equations, which could hamper the employment of the Newton-Raphson method. 

Such a problem does not arise in the Monte Carlo technique, since one solution is automatically picked up out of infinitely many possibilities by a dice. And that is sufficient indeed for our 
purpose because physical quantities are identical for any choice. Such an approach is not special and is indeed employed in other fields such as nuclear physics, in which deformed nuclei are
constructed in a way that is analogous to the one we adopt in this paper.  

The actual minimization procedure is the following. Given an initial configuration, we move one of the nodes in the radial direction slightly and see if this shift lowers the 
value of the energy function or not. If not, we cancel the shift. The amount of dislocation is a random variable but is limited to $\sim 1 \%$ of the lengths of the edges attached 
to the node. We do the same things in the lateral direction and sweep the entire mesh. We repeat this process until the value of the energy function is no longer lowered by a
certain amount ($\sim 1\%$). In the current version of the formulation, the connections of nodes by edges, i.e. the adjacency matrix in the graph-theoretical terms, are unchanged through the 
minimization process. This may need reconsideration, however, particularly when rotation is very rapid and resultant rotational equilibria are substantially deformed from the reference configuration.

Positions of the nodes on the outer boundary, which are referred to as the anchor nodes and have a vanishing mass, are fixed in this process, since an expansion in the radial direction would 
continue for ever otherwise. In some cases, however, the outer boundary comes too close to or goes too far from the nearby active nodes. We then shift the anchor nodes appropriately and fix them thereafter. 
We find that large deformations of triangular cells degrade accuracy of the energy function and in some cases generate a fictitious local minimum. To avoid such 
artifacts we apply a smoothing to the deformed portion of the grid whenever we detect too acute ($< 10^\circ$) an angle in triangles. 

\section{Applications} 
\begin{figure*}
\includegraphics[width=33pc]{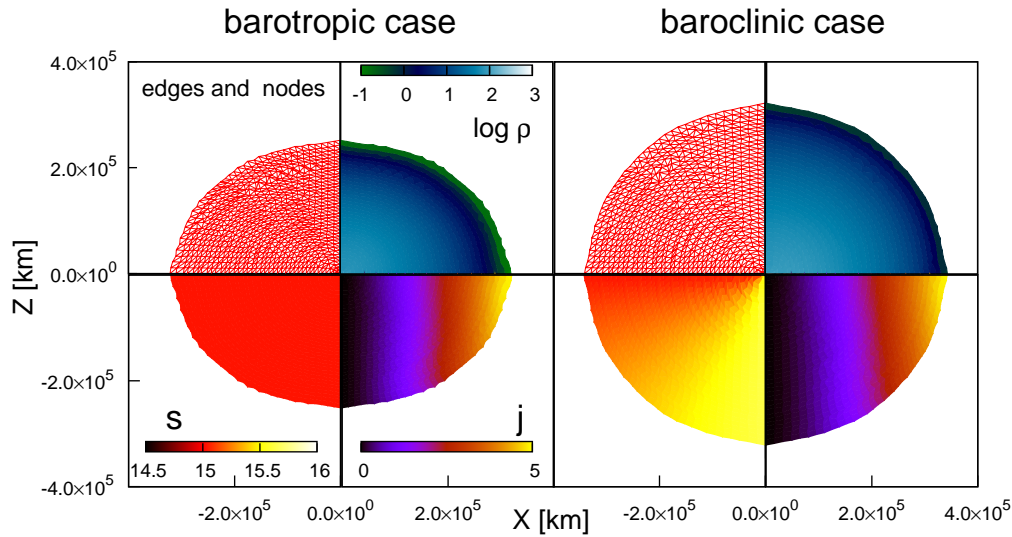}
\caption{\label{fig:01} (color on line). 
Structures of a star in rotational equilibria for barotropic (left four panels) and baroclinic (right four panels) EOS's. The upper left quadrants show the nodes and edges in the triangulated mesh. The other panels display clockwise the color contours of logarithmic density in ${\rm g/cm^3}$, specific entropy in ${\rm k_B}$ and specific angular momentum in 
$10^{18} {\rm cm^2/s}$. The color scales are identical for both cases.} 
\end{figure*}

We apply our new method to toy models of rotating stars. 
They are highly artificial admittedly and meant for a proof of principle alone. More realistic models and comparisons with existing models will be presented in the forthcoming full paper.
We assume that the stars are composed of hydrogens alone and employ an ideal-gas EOS. 
We first consider an isentropic model as a representative barotropic case. We assign an identical specific entropy to all nodes, $s_i=s_0$ with $s_0=15.0{\rm k_B}$. Here ${\rm k_B}$
is the Boltzmann's constant. The reference configuration is a spherical star of $0.60 {\rm M_\odot}$ with ${\rm M_\odot}$ being the solar mass. It is obtained by solving the Lane-Emden 
equation for the same specific entropy and mapping the result on a triangulated mesh with 852 nodes. The mass on each node $m_i$ is obtained from the density and 
volume assigned to the node in the initial configuration. As for the angular momentum, we assume in this spherically symmetric reference configuration an angular velocity distribution of 
the following form: $\Omega(\bvector{r}) = \Omega_0 \varpi_0^2 / (\varpi^2 + \varpi_0^2)$. Here $\varpi_0$ and $\Omega_0$ are model parameters and the former is set to the radius of the spherical 
star whereas the latter is $\Omega_0 = 10^{-6} \pi \times 5000$~rad/s, which is roughly 5000 times the solar angular velocity.
This gives the final angular velocity which is roughly half the critical angular velocity in the Roche approximation. Note that our models take into account self-gravity self-consistently. 

In this model, the rotational equilibrium is obtained after 525 sweeps when the value of the energy function does not change by more than 1\% in the last 50 sweeps. 
The results are given in the left four panels of FIG.~\ref{fig:01}. In the upper left quadrant the mesh configuration is displayed by nodes and edges. The distributions of density 
and specific entropy are shown in the upper right and lower left quadrants, respectively. Since this is an isentropic model, the specific entropy is constant in the stellar interior. 
From these pictures, it is evident that the rotational equilibrium is oblate, which is realized by lateral motions of nodes in our formulation. In the lower right quadrant, we show 
the distribution of the specific angular momentum. It is clear that the cylindrical rotation is realized as it should be. It is emphasized that this is not a trivial thing. In fact, the rotation 
law is not known a priori in our formulation. The cylindrical rotation is {\it not assumed but obtained} as a result of computations. This is a strong vindication that our formulation works well indeed. 
Although they are not shown here, other rotational laws including a rigid rotation are assumed in the reference configuration and we find in all cases cylindrical rotations in the outcomes.
We have also checked that our result is consistent with the one obtained with the method by \citet{hachisu86} for barotropic cases,
and the detail of which will be presented in our forthcoming paper.

Now we proceed to a baroclinic case. As an initial configuration we employ not a spherical star by the barotropic equilibrium just obtained. This is meant to observe clearly
how each node moves as a response to the introduction of baroclinicity. It is also helpful in terms of the numerical cost. We have confirmed that the formulation works equally well 
for an initially spherical configuration. The baroclinicity is most easily introduced by the artificial modification of the entropy distribution. Here we do this in the following way. 
Since we are using the ideal-gas EOS and the specific entropy is proportional to $K=P/ \rho^{5/3}$, we modify the value of $K$ on each node according to the relation:
$K_i = K_0 (1+ \theta_i/\pi)$, in which $K_0$ is the value of $K$ that gives $s_0$ in the barotropic model, and $\theta_i$ is the latitude of node $i$. This entropy 
distribution is admittedly quite artificial and will be much different from reality and it should be understood that this is just for illustrative purposes. Note, however, that the resultant configuration is stable against convection by the H{\o}iland criterion. 

We are again successful to obtain a rotational equilibrium also for this model after 255 sweeps, at which point the energy function does not change by more than 
1\% in the last 50 sweeps. The results are displayed in right four panels of FIG.~\ref{fig:01}. The grid configuration, density, specific angular momentum, specific entropy are shown 
in each quadrant clockwise from the upper left corner. It is clear from these pictures that the baroclinic configuration is much less oblate compared with the barotropic counter part. 
This is due to the enhanced thermal pressure near the rotation axis, which leads to the expansion of the star in the direction parallel to the axis. 
In our formulation, such motions are confirmed indeed by the comparison of the coordinates of the corresponding nodes in the reference and final configurations. 
The baroclinicity of the model is recognized most easily by the comparison between the distributions of density and specific entropy, since the isopycnic surfaces should coincide 
with the iso-entropic surfaces in the barotropic model, which is certainly not the case in this model. The baroclinicity of the model is also reflected in the angular momentum distribution, 
which is obviously {\it not} cylindrical. 
The inclination of iso-angular-momentum surfaces against the rotation axis can be also understood by the shifts of nodes mentioned above.
Incidentally, this angular momentum distribution is consistent with the Bjerknes-Rosseland rule.

\begin{figure}
\begin{center}
\includegraphics[width=15pc]{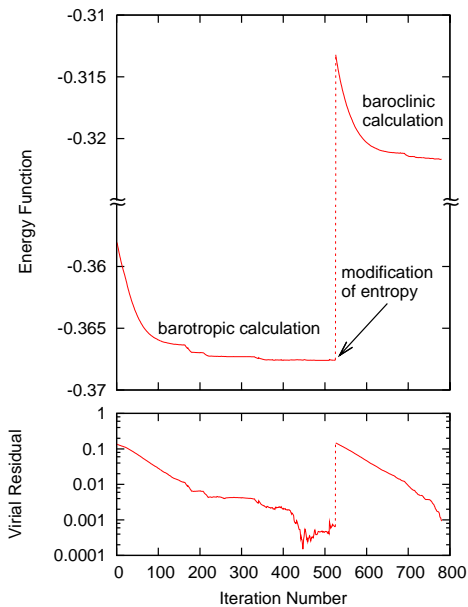}
\caption{\label{fig:02} 
The values of the energy function and of the virial residual in the minimization. The baroclinic case is attached at the end of the barotropic case as indicated by the arrow. 
The energy function is normalized by $G{\rm M_{\odot}}^2/{\rm R_{\odot}}$, in which ${\rm R_{\odot}}$ is the solar radius.
}
\end{center}
\end{figure}

In the upper panel of FIG.~\ref{fig:02} we show the histories of the values of the energy function in the minimizations for both the barotropic and baroclinic models. The result for the latter 
is attached at the end (the 525th iteration) of the result for the former. Except for a big jump at the junction point, the value of the energy function decreases most of the time. 
There are some small glitches occurring sporadically, however. They correspond to the smoothings, which are applied when a triangular cell is deformed too much and increase
the value of the energy function inevitably. These smoothings are necessary to ensure convergence to the true minimum, which is indeed observed in FIG.~\ref{fig:02}. Also shown 
in the lower panel is the residual in the virial relation defined as 
\begin{equation}
V_C =  \left |  \frac{2\,T+W+3\! \int \! P dV}{W} \right |,
\end{equation}
in which $T$, $W$ are the rotational and gravitational energies, respectively, 
and the third term is the volume integral of pressure. This is a measure of numerical accuracy commonly used in the literature. The results of both models are shown consecutively again. 
It is clear that $V_C \sim O(10^{-4})$ is achieved in both cases. It should be noted that this accuracy is attained with 852 nodes in this study.

\section{Discussions}

It is emphasized that the establishment of the formulation that can handle not only barotropic but also baroclinic rotational equilibria is itself a major achievement, since such attempts are not many in the literature~\citep{uryu94, espinoza13}. 

Note also that all previous 2D formulations are Eulerian, whereas our formulation is Lagrangian in 2D and we can trace the potentially complex movement of each fluid element during the quasi-static stellar evolutions. 
Its advantage in investigating the evolutionary series of rotating stars should be obvious.
Nuclear reactions occur locally in each fluid element and can be treated by network calculations on each node in our method.
Radiative energy transport will be handled by solving a diffusion equation on the triangulated mesh. Angular momentum transfer and mixing of elements by various (magneto) hydrodynamical
instabilities~\citep{maeder13} may be also approximated as diffusions on the Lagrangian coordinates just as in the current 1D stellar evolution models~\citep{ maeder00}. 

The meridional circulation, which is neglected in the current formulation, is another important issue\citep{LNP}. Since it is rather slow~\citep{espinoza13}, it is insignificant in force balance \citep{LNP}. It plays a major role, however, in the re-distribution of angular momentum and chemical abundance and, as a consequence, drives the evolution of rotating stars. Our strategy to treat this is the following: working on the time scale that is much longer than the dynamical time scale, 
we will incorporate this re-distribution of angular momentum and elements by
by solving the same transport equations on the triangulated mesh as employed in
the 1D models~\citep{LNP}.

The biggest challenge is the treatment of convectively unstable configurations. 
Whereas the energy functional is {\it minimized} in our formulation, convectively 
unstable equilibria do not correspond to 
the minimum but to a stationary point (or a saddle point) of the energy functional. A naive application of the formulation to such configurations will lead to a continued distortion of 
the mesh in the convection zone. A couple of ideas are currently being tested and will be reported elsewhere. 

As for the numerical techniques, the Monte Carlo method employed in this letter is rather slow and it takes days to get the minimization done at present. Although this may appear to be a 
concern in the application to the stellar evolution, it should be emphasized that the minimization is achieved in much shorter time if the initial configuration is very close to equilibrium, 
which is actually the case in the stellar evolution calculations. 

Appropriate triangulations, which are not touched in this letter, will become important in the application of the method to stars in advanced evolutionary stages, in which the stars
develop a core and mantles with large density contrasts. It may be necessary to extend the formulation so that it could handle multi-layers. 
Possible applications are not limited to these stars but will be extended to e.g. compact stars, proto-stars and planets. 

Incorporation of magnetic fields and/or general relativity should be considered in due course. 
Our new formulation will be a real break-through then.

\section*{Acknowledgements}
This work was partially supported by JSPS KAKENHI Grant Numbers 25105510, 24244036, 24103006.

\end{document}